# Novel Graph Processor Architecture, Prototype System, and Results

William S. Song, Vitaliy Gleyzer, Alexei Lomakin, Jeremy Kepner
MIT Lincoln Laboratory
Lexington, MA USA

*Abstract*— Graph algorithms are increasingly used in applications that exploit large databases. However, conventional processor architectures are inadequate for handling the throughput and memory requirements of graph computation. Lincoln Laboratory's graph-processor architecture represents a rethinking of parallel architectures for graph problems. Our processor utilizes innovations that include a sparse matrix-based graph instruction set, a cacheless memory system, accelerator-based architecture, a systolic sorter, high-bandwidth multi-dimensional toroidal communication network, and randomized communications. A field-programmable gate array (FPGA) prototype of the new graph processor has been developed with significant performance enhancement over conventional processors in graph computational throughput.

*Keywords—graph; processor; architecture; FPGA; prototype; results*

## I. Introduction

Many problems in computation and data analysis can be represented by graphs and analyzed using traditional graph analysis techniques. A graph, which is defined as a set of vertices connected by edges, as shown on the left in Figure 1, adapts well to presenting data and relationships. In general, graphs can also be represented as sparse matrices as shown in Figure 1 [1, 2] where an edge from vertex *i* to vertex *j* is represented as a matrix element in row *i* and column *j*.

Fig. 1.  Sparse matrix representation of graph.

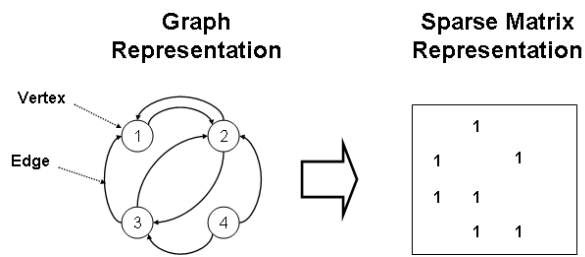

Increasingly, commercial and government applications are making use of these techniques [3] to analyze large databases that contain billions of vertices and edges requiring petabytes of storage capacity [4, 5, 6, 7, 8]. The datasets span a wide variety of application domains and contain (but not limited to) information from airborne sensor data, consumer purchasing patterns, financial market transactions, or bioinformatics data. For example, an analyst might be interested in spotting a cyber attack, elucidating a virus genealogy, or identifying a market niche.

For graph database applications, conventional processors perform poorly compared to non-graph applications because conventional processor architectures are generally not well matched to the flow of the graph computation. For example, most modern processors utilize cache-based memory in order to take advantage of highly localized memory access patterns. However, memory access patterns associated with graph processing are often random in nature and can result in high cache miss rates. In addition, graph algorithms incur significant computational overhead for index manipulation tasks required by graph traversing queries.

With the sparse-matrix-based graph representation, standard linear algebra matrix operations can be used to implement most graph algorithms. Furthermore, for benchmarking graph computation, sparse matrix operations can be used for estimating graph algorithm performance. Figure 2 shows an example of single-core computational throughput differences between dense-matrix processing, which has motivated most recent high-performance computing (HPC) processor architecture innovations, and sparse matrix graph processing [9, 10]. Shown in blue is a dense matrix-matrix multiply kernel running on single core PowerPC and Intel Xeon processors. In contrast, shown in red is a sparse matrix-matrix multiply kernel running on identical processors. As illustrated in the plot, the sparse matrix throughput is approximately 1000 times lower, which is consistent with typical performance seen by graph analysis applications.

Fig. 2.  Single core computational throughput differences between conventional and graph processing.

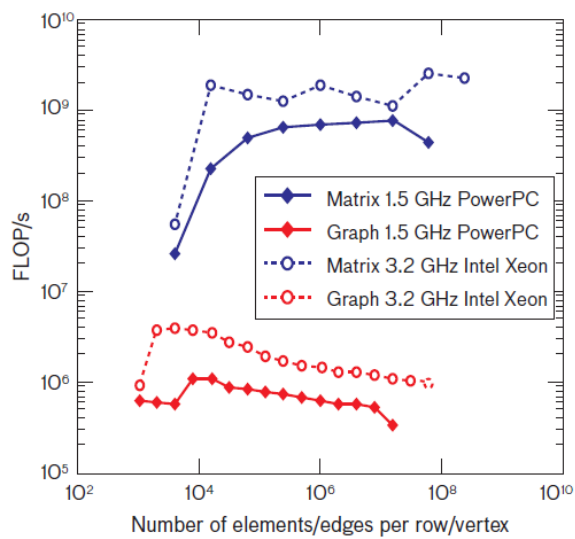

This material is based upon work supported by the Assistant Secretary of Defense for Research and Engineering under Air Force Contract No. FA8721-05-C-0002 and/or FA8702-15-D-0001. Any opinions, findings, conclusions or recommendations expressed in this material are those of the author(s) and do not necessarily reflect the views of the Assistant Secretary of Defense for Research and Engineering.

Parallel processors have often been used to speed up large conventional computing tasks. A parallel processor generally consists of multicore processors that are connected through a communication network, such that different portions of the computations can be done on different processors. For many scientific computing applications, these processors provide significant speedup over a single processor. However, large graph processing tasks often run inefficiently on conventional parallel processors. The speedup often levels off after only a small number of processors are utilized because the computing patterns for graph algorithms require much more communication between processor nodes than conventional, highly localized processing requires. The limited communication bandwidth of conventional parallel processors generally cannot keep pace with the demands of graph algorithms. In the past, numerous attempts have been made to speed up graph computations by optimizing processor architecture. Parallel processors such as Cray XMT and Thinking Machine's Connection Machine are example attempts to speed up large graph processing with specialized parallel architectures. However, inherent difficulties associated with graph processing, including distributed memory access, indices-related computation, and interprocessor communication, have limited the performance gains.

Lincoln Laboratory has been developing a promising new processor architecture that will deliver orders of magnitude higher computational throughput and power efficiency over the best commercial alternatives for large graph problems. The FPGA version of our processor is measured to be 10 times faster than conventional computers at the 100 Watt scale, and is projected to be over 100 times faster at the 1 Megawatt scale. Furthermore, the application-specific integrated circuit (ASIC) version of our processor is projected to be over 100 times faster at the 100 Watt scale and over 1000 times faster at the 1 Megawatt scale.

## II. GRAPH PROCESSOR

The new graph processor architecture represents a fundamental rethinking of the computer architecture for optimizing graph processing [9, 10, 13]. The instruction set is based on the emerging GraphBLAS standard [11, 12] that provides a common sparse matrix interface for graph analytics. The individual processor node— an architecture that is a great departure from the conventional von Neumann architecture— has local cacheless memory. All data computations, indices-related computations, and memory operations are handled by specialized accelerator modules rather than by the central processing unit (CPU). The processor nodes utilize new, efficient message-routing algorithms that are statistically optimized for communicating very small packets of data such as sparse matrix elements or partial products. The processor hardware design is also optimized for high-bandwidth six-dimensional (6D) toroidal communication network. Detailed analysis and simulations as well as a small-scale prototype have demonstrated up to several orders-of-magnitude increase in computational throughput and power efficiency for running complex graph algorithms on large distributed databases.

### A. Parallel Graph Processor Architecture Based on a Sparse Matrix Algebra Instruction Set

There are a number of advantages in implementing the graph algorithms as sparse matrix operations. One advantage is that the number of lines of code is significantly reduced in comparison to the amount of code required by traditional software that directly implements graph algorithms using conventional instruction sets. However, while this advantage can increase software development efficiency, it does not necessarily result in higher computational throughput in conventional processors.

Perhaps a more important advantage of implementing graph algorithms in sparse matrix operations is that it is much easier to design a parallel processor that computes sparse matrix operations rather than general graph algorithms. The instruction set can be vastly simplified because implementing sparse matrix–based graph algorithms requires surprisingly few base instructions. Another reason sparse matrix operations facilitate the designing of a processor architecture is that it is much easier to visualize the parallel computation and data movement of sparse matrix operations running on parallel processors. This advantage enables developers to come up with highly efficient architectures and hardware designs with less effort.

The graph processor is a highly specialized parallel processor optimized for distributed sparse matrix operations. The processor is targeted for implementing graph algorithms (converted to sparse matrix format) for analyzing large databases. Because large matrices do not fit into a single processor's memory and require more throughput than the single processor can provide, the approach is to distribute the large matrices over many processor nodes. Figure 4 shows the high-level architecture of the parallel processor. It consists of an array of specialized sparse matrix processors called node processors. The node processors are attached to the global communication network, and they are also attached to the global control processor through the global control bus.

Although the generic high-level architecture in Figure 3 appears quite similar to that of a conventional multiprocessor system, how it is implemented is significantly different from conventional parallel architecture implementations. One of the main differences is that the processor's instruction set is based on sparse matrix algebra operations [2] rather than on conventional instruction sets. Important instruction kernels include sparse matrix-matrix multiply, addition, subtraction, and division operations shown in Table 1. Individual element-level operators within these matrix operations, such as multiply and accumulate operators in the matrix-multiply operation, often need to be replaced with other arithmetic or logical operators, such as maximum, minimum, AND, OR, XOR, etc., in order to implement general graph algorithms. Numerous graph algorithms have already been converted to sparse matrix operations [2, 4, 5].

TABLE I. SPARSE MATRIX ALGEBRA-BASED PROCESSOR INSTRUCTION SET

| OPERATION | COMMENTS |
|---|---|
| C = A +.* B | Matrix-matrix multiply operation is the throughput driver for many important benchmark graph algorithms. Processor architecture is highly optimized for this operation. |
| C = A .± B <br> C = A .* B <br> C = A ./ B | Dot operations are performed within local memory. |
| B = op(k, A) | Operation with matrix and constant. This operation can also be used to redistribute matrix and sum columns or rows. |

Fig. 3. High level graph processor architecture.

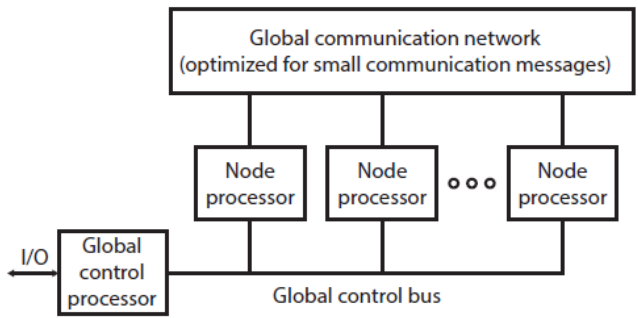

The other main differentiating feature of the new architecture is the high-bandwidth, low-power communication network that is tailored for communicating small messages. A typical message contains one matrix element or one partial product, which consists of the data value, row index, and column index. In contrast, a conventional communication network tries to maximize the message sizes in order to minimize the overhead associated with moving the data. A newly developed statistical routing algorithm with small message sizes greatly improves the communication throughput for graph processing. In addition, the bandwidth of the network hardware itself is significantly larger compared to the bandwidth of conventional parallel processors; this large bandwidth is needed to handle the demands of graph processing.

*B. Accelerator-Based Node Processor Architecture*

The architecture of the Laboratory's individual node processor is also a great departure from conventional cache-based von Neumann machines, which perform all computations in the CPU. This new architecture consists of a number of specialized modules, including matrix reader, matrix writer, sorter, arithmetic logic unit (ALU), and communication modules, as shown in Figure 4 [9, 10, 13]. The CPU is mainly used to provide the control and timing for the sparse matrix instructions. Most of the computation, communication, and memory operations are performed by the specialized modules that are designed to optimally perform the given tasks. There is no cache because the high cache miss rates slow down graph processing. In general, multiple modules are utilized simultaneously in performing sparse matrix computations.

The architecture based on the specialized accelerator module provides much higher computational throughput than the conventional von Neumann processor architecture by enabling highly parallel pipelined computations. In a conventional processor, the microprocessor is used to compute all the processing tasks, such as memory access, communication-related processing, arithmetic and logical operations, and control. These processing tasks are often done serially and take many clock cycles to perform, lowering the overall computational throughput. In the new architecture, these tasks are performed in parallel by separate specialized accelerator modules. These accelerator modules are designed for fast throughput using highly customized architectures. Ideally, they would be designed to keep up with the fastest data rate possible, which is processing one matrix element or one partial product within a single clock cycle in effective throughput. Further speedup may be gained by having multiple parallel versions of these modules to process multiple matrix elements or partial products per clock cycle.

The matrix reader and writer modules are designed to efficiently read and write the matrix data from the memory. The example formats include compressed sparse row (CSR), compressed sparse column (CSC), and coordinate (also called tuple) format (Figure 5). In the coordinate format, the data, row index, and column index are stored as triple pairs. In order to conserve storage space and improve lookup performance, in the CSR format, the element data and column index are stored as pairs in an array format. An additional array stores the row start address for each column so that these pointers can be used to look up the memory locations in which the rows are stored. The CSC format is similar except the columns are compressed instead of the rows.

Fig. 4. Node processor architecture utilizing accelator modules.

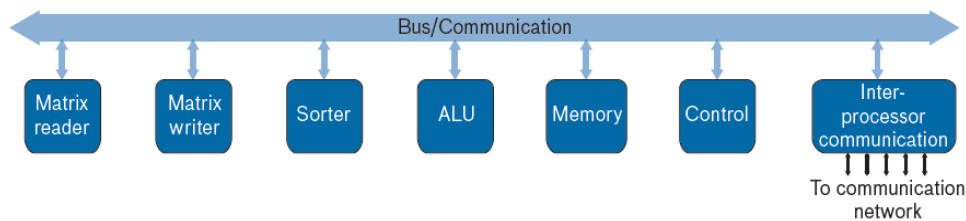

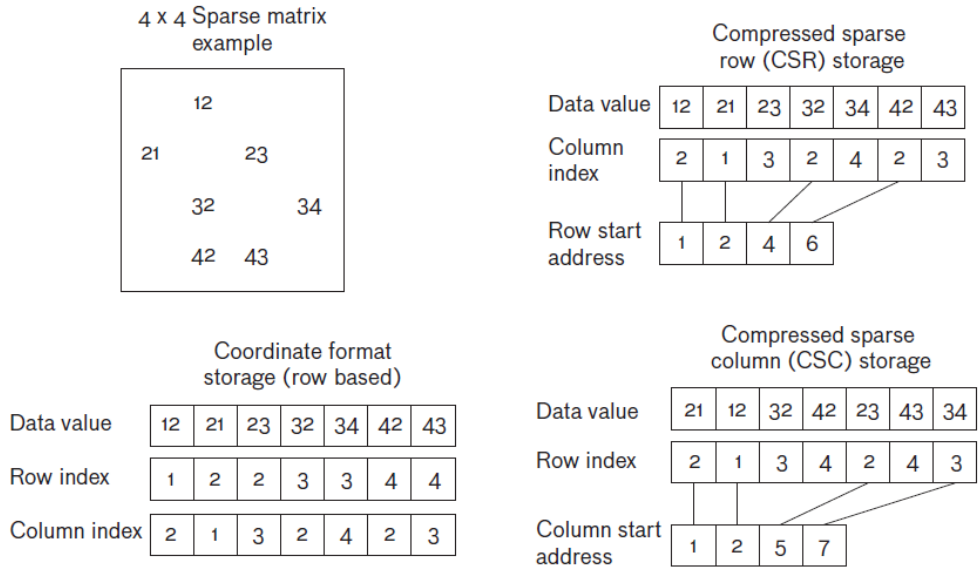

Fig. 5. Three formats for sparse matrix storage.

The matrix reader and writer modules are designed so that all the overhead operations—such as formatting matrix element data and indices for writing, generating pointer arrays for CSC and CSR for writing, and generating matrix element indices for reading—are performed automatically without requiring additional instructions. In this way, complexity associated with sparse matrix read and write operations is minimized, and memory interface operations are accelerated significantly.

The ALU module is designed to operate on the stream of sparse matrix elements or partial products instead of operating with a register file as in conventional processor architectures. The streaming method eliminates register load operations and increases the computational throughput. It generally performs designated arithmetic or logical operations on the data stream, depending on the indices. For example, the ALU module may accumulate successive matrix elements only if the element indices match exactly. Because these matrix operations perform computations only when the indices match, this feature is useful for sparse-matrix add and multiply operations.

The communications module handles the communication between processor nodes. It takes the matrix element or partial product and makes a communication message that includes the matrix element in coordinate format and a header that contains the destination processor address. The header may also contain error detection and correction bits and other relevant information, such as the priority of the message. The communication messages are then sent to the global communication network and are forwarded to the destination nodes. The communications module also decodes the received messages, performs error correction, and outputs the matrix element or partial product into the node in coordinate format.

The memory for the node processor can be implemented with various types of memory including static random-access memory (SRAM), dynamic RAM, and synchronous DRAM. Nonvolatile memory, such as Flash memory, may be used for long-term storage and for instances when the storage requirement is high.

The node controller module is responsible for setting up and coordinating the sparse matrix operations. Before a sparse matrix operation, the controller module loads the control variables into the control registers and control memory of the accelerator modules by using the local control bus. The control variables include types of sparse matrix operations to be performed, matrix memory storage locations, matrix distribution mapping, and other relevant information. The controller module also performs timing and control. The node controller module can be implemented with a conventional general-purpose microprocessor. This particular microprocessor may also have a cache since the processing is mostly conventional processing. The node controller can also perform other processing tasks that are not well supported by the accelerator modules, such as creating an identity matrix and checking to see if a matrix is empty across all processor nodes. The controller module is tied to the global control bus, which is used to load the data and programs to and from the nodes, and to perform the global computation process control.

The sorter module is used for sorting the matrix element indices for storage and for finding matching element indices during matrix operations. It is one of the most critical modules in graph processing because more than 95% of computational throughput can be associated with the sorting of indices. The sparse matrix and graph operations consist mainly of figuring out which element or partial product should be operated on. In contrast, relatively few actual element-level operations get performed. In order to meet the computational throughput requirement, the systolic $k$-way systolic merge sorter architecture [14] was developed to provide significantly higher throughput than the conventional merge sorters.

The $k$-way merge sorter sorts long sequences of numbers by using a recursive "divide and conquer" approach. It divides the sequence into $k$ sequences that have equal, or as equal as possible, lengths. The $k$ shorter sequences are then sorted independently and merged to produce the sorted result. The sorting of $k$ shorter sequences can also be divided into $k$ even shorter sequences and sorted recursively by using the same merge sort algorithm. This process is recursively repeated until the divided sequence length reaches 1. The sorting process takes order $n\log_k n$ memory cycles. The $k$-way merge sort is $\log_2 k$ times faster than the 2-way merge sort process when $k$ is greater than 2. For example, when $k = 32$, the $k$-way merge sorter has five times greater sorter throughput than the 2-way merge sorter. The main difficulty with implementing a $k$-way merge sorter in a conventional processor is that it takes many clock cycles to figure out the smallest (or largest) value among $k$ entries during each step of the merge sorting process. Ideally, the smallest value of $k$ should be computed within one processor clock cycle for the maximum sorter throughput. The 100% efficient systolic merge sorter [9] can achieve this performance requirement using $k$ linear systolic array cells and it is particularly well suited for FPGA and integrated circuit (IC) implementation since it consists of repeated systolic cells with nearest-neighbor-only communications.

### C. 6D Toroidal Communication Network and Randomized Message Routing

The new graph processor architecture is a parallel processor interconnected in a 6D toroidal configuration using high bandwidth optical links. The 6D toroid provides much higher communication performance than lower-dimensional toroids because of the higher bisection bandwidth.

The communication network is designed as a packet-routing network optimized to support small packet sizes that are as small as a single sparse matrix element. The network scheduling and protocol are designed such that successive communication packets from a node would have randomized destinations in order to minimize network congestion [15]. This design is a great contrast to typical conventional multiprocessor message-routing schemes that are based on much larger message sizes and globally arbitrated routing that are used in order to minimize the message-routing overhead. However, large message-based communications are often difficult to route and can have a relatively high message contention rate caused by the long time periods during which the involved communication links are tied up. The small message sizes, along with randomized destination routing, minimize message contentions and improve the overall network communication throughput. Figure 6 shows the 512-node (8 × 8 × 8) 3D toroidal network (drawn as 3 × 3 × 3 network for illustration purposes) simulation based on randomized destination communication versus unique destination communication. Even though both routing methods are based on small message sizes, the unique destination routing has a message contention rate that is closer to the contention rate of conventional routing algorithms that are based on large message sizes. The randomized destination routing achieved approximately six times higher data rate and network utilization efficiency in the simulation using an identical network.

Fig. 6. Randomized destination vs. unique destination packet communication.

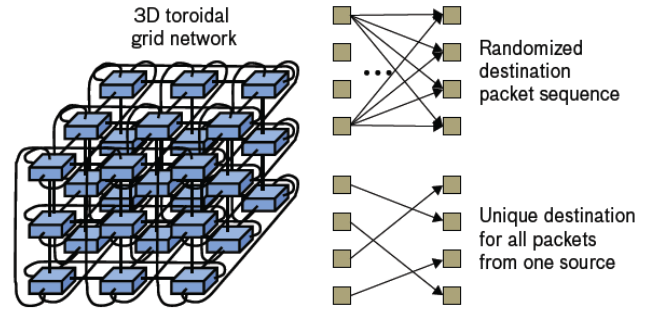

### III. FPGA PROTOTYPE DEVELOPMENT AND PERFORMANCE MEASUREMENT

Lincoln Laboratory has developed an FPGA prototype of the graph processor using commercial FPGA boards as shown in Figure 7. Each board has one large FPGA and two 4-GByte DDR3 memory banks. Two graph processor nodes are implemented in each board. A small 4-board chassis implements an 8-node graph processor tied together with 1D toroidal network. Since the commercial board offered limited scalability due to limited number of communication ports for network connection, the larger prototypes will be developed in the future using custom FPGA boards that can support 6D toroidal network and up to 1 million nodes.

Fig. 7. FPGA prototype of the graph processor.

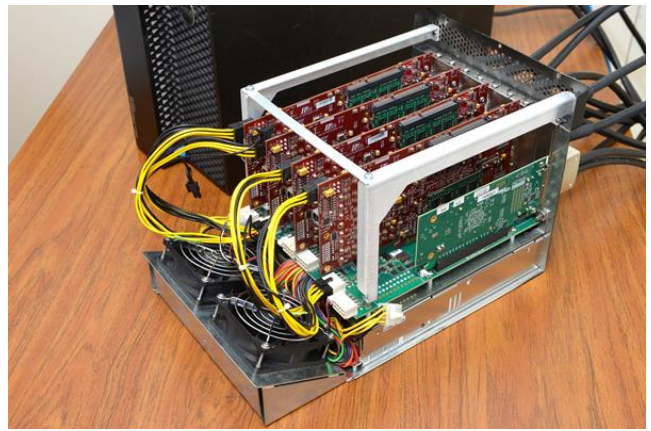

Fig. 8. Graph processor prototype performance measurements and projections.

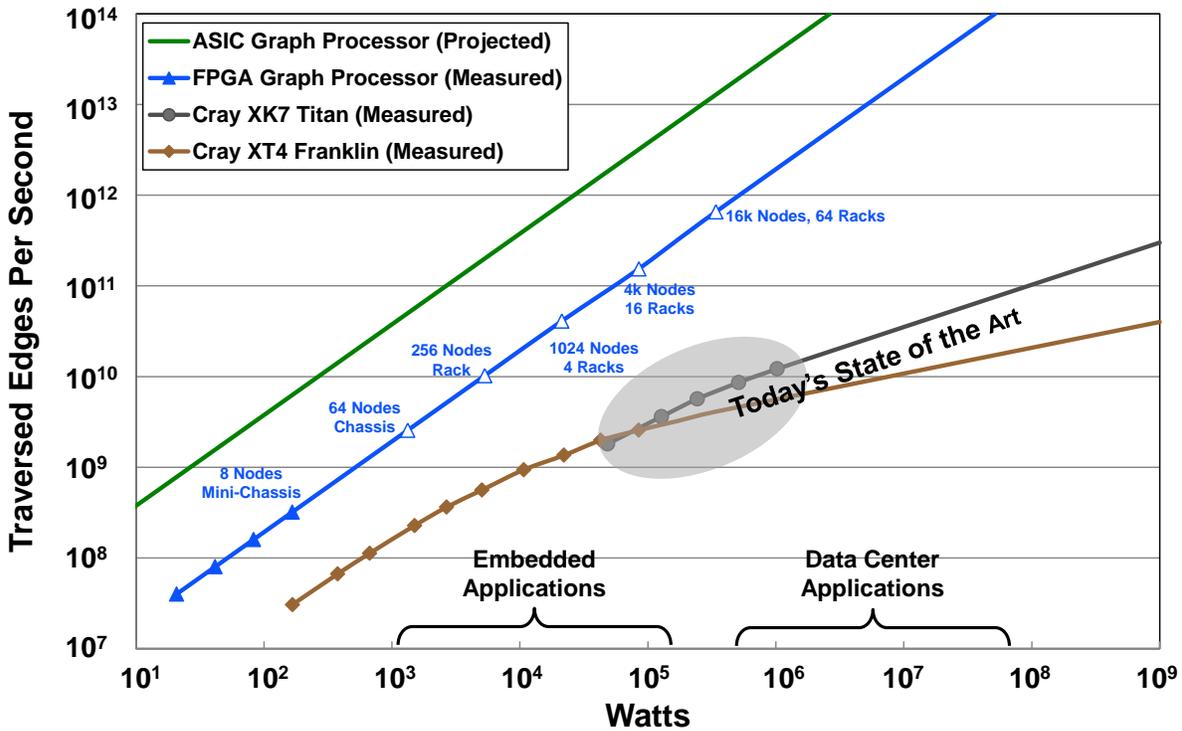

The performance of the FPGA prototype processor was measured using sparse matrix-matrix multiply operations on power law matrices. Figure 8 shows the measured (1, 2, 4, and 8 nodes) and projected (>8 nodes) performance in terms of traversed edges per second vs. power consumption of the prototype processors with various number of nodes. For projected performance, a detailed simulation of the architecture was performed and bit-level accurate simulation models were used to simulate up to 1024-node processor running sparse matrix-matrix multiply kernels. The projected and measured performances were identical for the prototype processors with 8 or less nodes. Also plotted are performances of Cray XK7 and XT4 supercomputer systems for comparison. [16, 17]

The FPGA prototype processor has an order of magnitude higher measured power efficiency compared to conventional processors at small number of nodes. At higher number of nodes, projected power efficiency difference increases to several orders of magnitude due to the linear speedup of computational throughput. There are multiple features of the architecture enabling this linear speedup. One is the sufficient communication bandwidth between processor nodes enabled by the high-bandwidth network and the advanced randomized packet routing algorithm. Another factor is the highly-efficient load balancing algorithm. For the graph processor to run efficiently, it needs to balance the memory usage and processing among all the nodes. That means the number of sparse matrix elements stored, partial products generated, and partial products accumulated in the processor nodes have to be well balanced with other nodes. Numerous advanced algorithms were developed to ensure optimal memory and processing load balancing. [9, 10, 13]

In the future, even higher throughput performance and power efficiency can be gained by using ASICs instead of FPGAs, as shown with the green performance line in Figure 8. Much higher processor circuit density and power efficiency can be enabled by ASIC. Memory performance and power efficiency can be increased by going from DDR3 SDRAMs to DDR4 SDRAMs. For optical communications, utilizing wavelength division multiplexing (WDM) low-power silicon optic technology can significantly increase communication bit rate per fiber and reduce communications power consumption [18, 19].

IV. CONCLUSION

MIT Lincoln Laboratory's graph-processor architecture represents a fundamental rethinking of architectures for graphs. It utilizes multiple innovations to overcome shortcomings of conventional architectures, including: a sparse matrix-based graph instruction set, an accelerator-based architecture, a high-performance systolic sorter, randomized communications, a cache-less memory system, and a high-bandwidth multi-dimensional communication network.

MIT Lincoln Laboratory has developed a graph-processor prototype based on commercial FPGA technology. The sparse matrix-matrix multiply, which is one of the most important kernels in graph analytics, has been demonstrated on an 8-node prototype processor and the work is under way to develop much larger prototype systems. The current small-scale prototype is designed to demonstrate both the scalability of the architecture as well as unprecedented graph algorithm throughput and power efficiency. The measured power efficiency at small number of nodes is an order of magnitude higher than commercial processors, and up to several orders of magnitude higher performance is projected at higher node counts. In the future, this implementation can be further optimized using a custom ASIC implementation, further enhancing computational throughput and power efficiency.

References

[1] J. Kepner and J. Gilbert, Graph Algorithms in the Language of Linear Algebra. Philadelphia: SIAM Press, 2011.

[2] S.H. Roosta, Parallel Processing and Parallel Algorithms, Theory and Computation. New York: Springer-Verlag, 2000.

[3] T.H. Cormen, C.E. Leiserson, R.L. Rivest, and C. Stein, Introduction to Algorithms. Cambridge, Mass.: The MIT Press, 2001

[4] D.A. Bader, "Exascale Analytics for Massive Social Networks,'" Minisymposium on High-Performance Computing on Massive Real-World Graphs, 2009 SIAM Annual Meeting (AN09), Denver, CO, July 6-10, 2009.

[5] J. Kepner, D. Bader & J. Gilbert, "Massive Graphs: Big Compute meets Big Data," Minisymposium at SIAM Annual Meeting, July 9-13, 2012, Minneapolis, Minnesota

[6] P. Burkhardt, C. Waring, "A cloud-based approach to big graphs," IEEE High Performance Extreme Computing Conference (HPEC), Sep. 2015.

[7] C. Wickramaarachchi, M. Frincuy, P. Small, and V. Prasanna, "Fast Parallel Algorithm For Unfolding Of Communities In Large Graphs," IEEE High Performance Extreme Computing Conference (HPEC), Sep. 2015.

[8] A. Kumbhare, M. Frincu, C. Raghavendra, and V. Prasanna, "Efficient Extraction of High Centrality Vertices in Distributed Graphs," IEEE High Performance Extreme Computing Conference (HPEC), Sep. 2014.

[9] W.S. Song, J. Kepner, H.T. Nguyen, J.I. Kramer, V. Gleyzer, J.R. Mann, A.H. Horst, L.L. Retherford, R.A. Bond, N.T. Bliss, E.I. Robinson, S. Mohindra, and J. Mullen, "3-D Graph Processor," Workshop on High Performance Embedded Computing, September 2010, available at http://www.ll.mit.edu/ HPEC/agendas/proc10/agenda.html.

[10] W.S. Song, J. Kepner, V. Gleyzer, H.T. Nguyen, and J.I. Kramer, "Novel Graph Processor Architecture," Massachusetts Institute of Technology Lincoln Laboratory Journal, vol. 20, no. 1, pp. 92-104, 2013.

[11] T. Mattson, D. Bader, J. Berry, A. Buluc, J. Dongarra, C. Faloutsos, J. Feo, J. Gilbert, J. Gonzalez, B. Hendrickson, J. Kepner, C. Leiserson, A. Lumsdaine, D. Padua, S. Poole, S. Reinhardt, M. Stonebraker, S. Wallach, & A. Yoo, Standards for Graph Algorithm Primitives, IEEE High Performance Extreme Computing (HPEC), Sep 2013

[12] M. Wolf, J. Berry, and D. Stark, "A task-based linear algebra Building Blocks approach for scalable graph analytics," IEEE High Performance Extreme Computing Conference (HPEC), Sep. 2015.

[13] W.S. Song, "Processor for large Graph Algorithm Computations and Matrix Operations," US Patent No. 8,751,556, June 10, 2014.

[14] William S. Song, "Systolic Merge Sorter," US Patent No. 8,190,943, May 29, 2012.

[15] W.S. Song, "Microprocessor Communication Networks," US Patent No. 8,819,272, August 26, 2014.

[16] A. Buluc and J. Gilbert. "Parallel Sparse Matrix-Matrix Multiplication and Indexing: Implementation and Experiments,".SIAM Journal of Scientific Computing (SISC), 34(4):170-191, 2012.

[17] A. Azad, G. Ballard, A. Buluc, J. Demmel, L. Grigori, O. Schwartz, S. Toledo, and S. Williams, "Exploiting Multiple Levels of Parallelism in Sparse Matrix-Matrix Multiplication," Technical Report 1510.00844.arXiv.

[18] Y. Vlasov, "Silicon CMOS-integrated nano-photonics for computer and data communications beyond 100G," IEEE Communications Magazine, vol. 50, issue 2, pp. S67-572, Feb. 2012.

[19] G. Reed, "Silicon Photonics: The State of The Art," Wiley Online Library, John Wiley & Sons, Ltd., 2008.